\documentclass[5p,times]{elsarticle}

\usepackage[figuresright]{rotating}

\usepackage[ruled,vlined]{algorithm2e}

\usepackage[most]{tcolorbox}
\tcbuselibrary{skins,breakable}
\usepackage{xcolor}
\usepackage{ragged2e}
\usepackage{setspace}
\usepackage[inline]{enumitem}
\usepackage{caption}

\newcommand{\pvar}[1]{\mbox{\textcolor{blue!60!black}{\textsf{\{#1\}}}}}

\newtcolorbox{PromptCard}[2][]{%
  enhanced,
  breakable,
  colback=black!1,
  colframe=blue!45!black,
  boxrule=0.6pt,
  arc=2.2mm,
  left=4mm,right=4mm,top=2.6mm,bottom=2.6mm,
  boxsep=1mm,
  colbacktitle=blue!45!black,
  coltitle=white,
  fonttitle=\bfseries\small,
  title={#2},
  attach boxed title to top left={xshift=2mm,yshift=-1.0mm},
  boxed title style={boxrule=0pt,arc=2mm,left=3.2mm,right=3.2mm,top=0.8mm,bottom=0.8mm},
  fontupper=\small\RaggedRight\setstretch{1.06},
  #1
}

\usepackage{amsmath,amssymb,capt-of}
\usepackage{amssymb}  
\usepackage{booktabs} 
\usepackage{pifont}
\newcommand{\xmark}{\ding{55}} 

\usepackage{wasysym}
\usepackage[utf8]{inputenc}
\usepackage[super]{nth}

\usepackage[table]{xcolor}

\usepackage{graphicx,booktabs,tabularx,array}
\newcolumntype{C}[1]{>{\centering\arraybackslash}p{#1}}
\newcolumntype{L}[1]{>{\raggedright\arraybackslash}p{#1}}

\usepackage{listings}

\usepackage{multirow}
\usepackage{hhline}
\usepackage{colortbl}

\usepackage[acronym,toc]{glossaries}

\usepackage[inline]{enumitem}

\usepackage{array}

\usepackage{amsmath}

\usepackage{caption}
\usepackage{subcaption}
\usepackage{float}

\usepackage{lastpage,fancyhdr,graphicx}

\usepackage{multirow}
\usepackage{tabularx}

\usepackage{etoolbox}
\makeatletter
\patchcmd{\@verbatim}
  {\verbatim@font}
  {\verbatim@font\small}
  {}{}
\makeatother

\usepackage{makecell}

\usepackage{listings}
\usepackage{pifont}
\definecolor{ao}{rgb}{0.0, 0.5, 0.0}
\definecolor{amber}{rgb}{1.0, 0.49, 0.0}

\usepackage{nameref,hyperref}

\journal{Information Sciences}

\begin{document}
\begin{frontmatter}


\title{Agent-based simulation of online social networks and disinformation}

\author[addr1]{Alejandro~Buitrago~L\'opez}\texorpdfstring{\corref{cor1}}{}
\ead{alejandro.buitragol@um.es}

\author[addr1]{Alberto~Ortega~Pastor}
\ead{alberto.o.p@um.es}

\author[addr1]{David~Montoro~Aguilera}
\ead{d.montoroaguilera@um.es}

\author[addr1]{Mario~Fern\'andez~T\'arraga}
\ead{mario.fernandezt@um.es}

\author[addr1]{Jes\'us Verd\'u~Chac\'on}
\ead{jesus.v.c@um.es}

\author[addr1]{Javier~Pastor-Galindo}
\ead{javierpgo@um.es}

\author[addr1]{Jos\'e~A.~Ruip\'erez-Valiente}
\ead{jruiperez@um.es}

\cortext[cor1]{Corresponding author}
\address[addr1]{Department of Information and Communications Engineering, University of Murcia, 30100 Murcia, Spain}

\begin{abstract}
Research on online social networks (OSNs) is often hindered by platform opacity, limited access to data, and ethical constraints. Simulation offer a valuable alternative, but existing frameworks frequently lack realism and explainability. This paper presents a simulation framework that models synthetic social networks with agents endowed with demographic-based personality traits and finite-state behavioral automata, enabling realistic and interpretable actions. A generative module powered by a large language model (LLM) produces context-aware social media posts consistent with each agent's profile and memory. In parallel, a red module implements DISARM-inspired workflows to orchestrate disinformation campaigns executed by malicious agents targeting simulated audiences. A Mastodon-based visualization layer supports real-time inspection and post-hoc validation of agent activity within a familiar interface. We evaluate the resulting synthetic social networks using topological metrics and LLM-based content assessments, demonstrating structural, behavioral, and linguistic realism. Overall, the framework enables the creation of customizable and controllable social network environments for studying information dynamics and the effects of disinformation.
\end{abstract}


\end{frontmatter}

\section{Introduction}\label{introduction}
In recent years, online social networks (OSNs) have attracted increasing attention as a means to study patterns of human interaction, information spread, and the formation of collective behaviors~\cite{Conte2012}. These platforms have reshaped how opinions are formed, forged alliances, and amplified conflicts in civic and geopolitical contexts. Beyond their role in fostering social connectivity, OSNs have also become critical areas for disinformation and coordinated influence operations that manipulate public opinion and disrupt democratic processes~\cite{bennett2018disinformation}. To counter these threats, often orchestrated by state and nonstate actors~\cite{theocharidou2021enisa}, it is worth understanding how information and influence propagate through digital environments, but studying such processes empirically remains challenging.

Applied research has demonstrated the impact of social media in information dissemination \cite{doi:10.1126/science.aap9559}, political polarization~\cite{hateSpeechAlex}, and collective behavior \cite{pastorgalindo2025}. However, fundamental gaps persist in our ability to respond to research questions experimentally, in configurable environments. The inherent opacity of platform algorithms, limited access to granular user data, and ethical constraints on intervening in real-world systems significantly hinder our capacity to test hypotheses or evaluate countermeasures in controlled settings~\cite{Ferraro2024}. This limitation is particularly acute when studying adversarial behaviors, such as disinformation campaigns or coordinated inauthentic activity~\cite{Terp2022}, where ground truth is scarce, experimentation is infeasible, and consequences are often politically sensitive.

In response to these limitations, simulation poses a promising alternative for customized experimentation, offering a safe and controllable sandbox to explore what cannot be directly observed or tested in live systems \cite{oasis}. Crucially, simulations can replicate not only organic user behaviors but also adversarial strategies\cite{hu2025simulatingrumorspreadingsocial}. Yet, the effectiveness of such simulations depends on critical factors such as the structural fidelity of the synthetic OSN, the behavioral realism of its constituent users, and the effectiveness of influence strategies. In fact, most existing simulation frameworks simplify agent behavior through static rules and model interactions as abstract diffusion processes, neglecting the cognitive and communicative complexity of real-world social dynamics \cite{surveyAlex}. Nevertheless, recent advances in Large Language Models (LLMs) provide a promising opportunity to enhance social experimentation. 

LLMs can be combined with symbolic behavior models and memory structures to enable agents to dynamically replicate human behavior~\cite{zhang2024surveymemorymechanismlarge}, thereby enhancing the realism of social simulations \cite{10492674}. For instance, human-like routines and interactions in small town environment simulations~\cite{park2023generativeagentsinteractivesimulacra}, agents' simulations in video games \cite{wang2023voyageropenendedembodiedagent}, and recommender systems \cite{recommender}, among others \cite{mou2024individualsocietysurveysocial}. Particularly for agent-based social networks, LLMs enable the production of user-generated content and the simulation of information spreading \cite{oasis}. Moreover, some works reproduce malicious accounts with deceptive behavior \cite{xu2025forewarnedforearmedsurveylarge} and disseminate fake news~\cite {Terp2022}.

However, key gaps remain in generative agent-based social networks. Frameworks such as AgentSociety~\cite{piao2025agentsocietylargescalesimulationllmdriven} and OASIS~\cite{yang2025oasisopenagentsocial} succeed in scaling to thousands or even millions of agents, but offer limited control over fine-grained behavioral routines, which constrains reproducibility and interpretability. Content realism is also uneven: platforms like Chirper.ai~\cite{zhu2025characterizingllmdrivensocialnetwork} and Social Simulacra~\cite{10.1145/3526113.3545616} prioritize linguistic plausibility but often lack narrative coherence or agenda persistence across posts and time. Finally, adversarial settings remain underdeveloped. BotSim~\cite{qiao2024botsimllmpoweredmalicioussocial} and Y Social~\cite{rossetti2024ysocialllmpoweredsocial} demonstrate that LLMs can generate manipulative or persuasive content, but they rarely capture the structured tactics, techniques, and procedures (TTPs) that characterize coordinated campaigns in real OSNs~\cite{hu2025simulatingrumorspreadingsocial}. As a result, current frameworks fall short of modeling the evolving, intentional dynamics of influence operations and the cumulative patterns of engagement and manipulation over time.

To address the identified gaps, this work introduces a simulation framework that combines realistic network generation, a generative module capable of generating content, and adversarial behavior modeling. In particular, this work introduces a modular framework that unifies the structural, behavioral, and adversarial dimensions of online social networks. Unlike prior approaches that treat LLM agents as black-box generators or neglect adversarial realism, our system provides a configurable and interpretable testbed. At the structural level, it generates synthetic OSNs that reproduce key properties of real platforms through homophily, semantic similarity, and triadic closure. At the behavioral level, it equips agents with finite-state decision models, memory, and LLM-based content generation, producing context-aware and linguistically realistic interactions. At the adversarial level, it operationalizes disinformation campaigns via standardized workflows inspired by DISARM, enabling the programmable deployment of tactics such as narrative injection, amplification, and audience targeting.

Altogether, the framework establishes the first simulation environment that unifies structural realism, interpretable generative agents, and programmable disinformation workflows, offering a reproducible testbed for both scientific analysis of emergent OSN phenomena and systematic evaluation of countermeasures. The remainder of the paper is organized as follows. Section~\ref{sec:sota} presents related work. Section~\ref{sec:overview} introduces the architecture and core components of the proposed framework. Section~\ref{sec:simulation} details the simulation process. Section~\ref{sec:stack} describes the generative agent stack. Section~\ref{sec:deployment} focuses on deploying synthetic disinformation strategies. Section~\ref{sec:results} reports the experimental results across different simulation scales. Section~\ref{sec:visualization} focuses on the visualization features as a primary example of third-party extensibility. Finally, Section~\ref{sec:conclusion} concludes the paper and outlines directions for future work.

\section{Related works}\label{sec:sota}

Simulation has long been used to investigate the structural and dynamic properties of OSNs, offering a means to model large-scale interactions under controlled conditions. Traditional approaches have emphasized rule-based agents and abstract diffusion mechanisms, often overlooking the cognitive and communicative complexity of real users~\cite {surveyAlex}. In this section, we focus on social simulations and disinformation phenomena using generative AI and LLMs. 

A set of works discusses the opportunities and challenges that generative AI provides to social simulations \cite{10492674}. Specifically, Papachristou et al.~\cite{papachristou2024networkformationdynamicsmultillms} showed that LLMs like GPT and Claude can reproduce network formation patterns like homophily and triadic closure, yielding small-world structures. In parallel, Anthis et al.~\cite{anthis2025llmsocialsimulationspromising} discuss methodological limitations of LLM-based simulations and propose best practices to enhance their use as surrogates for human behavior. Complementarily, Ashery et al.~\cite{doi:10.1126/sciadv.adu9368} explore how decentralized LLM populations can develop emergent social conventions and collective biases through repeated interaction, highlighting the potential for norm formation even in the absence of explicit network structure.

In a related line of research, BotSim~\cite{qiao2024botsimllmpoweredmalicioussocial} examines the ability of LLM-based agents to engage in manipulative behavior during dyadic interactions. Their study models one-on-one conversations where an agent attempts to influence a partner's belief through deception, emotional appeal, or misleading reasoning.  While this work does not simulate network-level interactions, it explores the role of LLMs as potential vectors of manipulation.

Other approaches integrate LLMs into more holistic simulations of societal dynamics. AgentSociety~\cite{piao2025agentsocietylargescalesimulationllmdriven} presents a large-scale platform that combines LLM-based agents with a simulated environment encompassing mobility, communication, and economic activity. Enabling over $10,000$ agents to simulate polarization, message diffusion, policy experiments, and disasters. While AgentSociety captures macro patterns, relying on LLMs as opaque generators limits interpretability and agent-level control.

A complementary direction focuses on using LLM-driven agents to simulate entire OSNs. OASIS \cite{yang2025oasisopenagentsocial} introduced a framework for simulating large-scale OSNs. It models platforms like $\mathbb{X}$ using realistic user profiles, dynamic follower networks, and diverse action spaces, including posting, commenting, and following. It incorporates a recommendation system and a temporal activity engine to reflect platform-specific dynamics and user behaviors over time. The system can simulate up to one million agents and reproduces key social phenomena such as information propagation, group polarization, and herd effects. Nevertheless, OASIS relies primarily on emergent behaviors from LLM agents interacting within platform-specific constraints.

In this vein, MOSAIC~\cite{liu2025mosaicmodelingsocialai}, proposed an open-source OSN simulator where LLM-driven agents interact through posting, sharing, and moderation actions within a directed social graph. The framework integrates memory, reflection, and diverse moderation modes (community-based, third-party, and hybrid), enabling the evaluation of content regulation policies. Despite its robust instrumentation and empirical validation, MOSAIC operates at limited scale and exhibits a gap between agents' stated intentions and aggregate behaviors, constraining its cognitive realism.

\begin{table*}[t]
\centering
\begin{tabular}{lccccccc}
\toprule
\textbf{Work} & 
\rotatebox{45}{\textbf{OSN simulation}} & 
\rotatebox{45}{\textbf{LLM-powered agents}} & 
\rotatebox{45}{\textbf{Assessment}} & 
\rotatebox{45}{\textbf{Agent memory}} & 
\rotatebox{45}{\textbf{Behavioral control}} & 
\rotatebox{45}{\textbf{Red agents}} & 
\rotatebox{45}{\textbf{Multimodality}} \\
\midrule
Network formation \cite{papachristou2024networkformationdynamicsmultillms} & \xmark & \checkmark & \checkmark & \xmark & \xmark & \xmark & \xmark \\
AgentSociety \cite{piao2025agentsocietylargescalesimulationllmdriven} & \xmark & \checkmark & \checkmark & \checkmark & \xmark & \xmark & \checkmark \\
Emergent bias \cite{doi:10.1126/sciadv.adu9368} & \xmark & \checkmark & \checkmark & \checkmark & \xmark & \xmark & \xmark \\
OASIS \cite{yang2025oasisopenagentsocial} & \checkmark & \checkmark & \xmark & \checkmark & \xmark & \xmark & \checkmark \\
Chirper.ai \cite{zhu2025characterizingllmdrivensocialnetwork} & \checkmark & \checkmark & \checkmark & \xmark & \xmark & \xmark & \xmark \\
Rumor spreading \cite{hu2025simulatingrumorspreadingsocial} & \checkmark & \checkmark & \xmark & \xmark & \xmark & \checkmark & \xmark \\
Influence dynamics \cite{nasim2025simulatinginfluencedynamicsllm} & \checkmark & \checkmark & \xmark & \xmark & \xmark & \checkmark & \xmark \\
BotSim \cite{qiao2024botsimllmpoweredmalicioussocial} & \xmark & \checkmark & \checkmark & \checkmark & \checkmark & \checkmark & \xmark \\
Y Social \cite{rossetti2024ysocialllmpoweredsocial} & \xmark & \checkmark & \checkmark & \checkmark & \xmark & \checkmark & \xmark \\
MOSAIC \cite{liu2025mosaicmodelingsocialai} & \checkmark & \checkmark & \checkmark & \checkmark & \checkmark & \xmark & \xmark \\
Social Simulacra \cite{10.1145/3526113.3545616} & \xmark & \checkmark & \checkmark & \xmark & \xmark & \xmark & \xmark \\
Oppi \cite{oppi2024} & \xmark & \xmark & \checkmark & \xmark & \xmark & \checkmark & \checkmark \\
\textbf{Our} & \checkmark & \checkmark & \checkmark & \checkmark & \checkmark & \checkmark & \xmark \\
\bottomrule
\end{tabular}
\caption{Comparative overview of recent simulation frameworks along seven dimensions: OSN simulation, use of LLM-powered agents, empirical assessment or validation, agent memory modeling, behavioral control mechanisms, inclusion of adversarial RED agents, and support for multimodality (e.g., text–image or cross-modal content).}
\label{tab:comparison}
\end{table*}

In a related direction, Chirper.ai \cite{zhu2025characterizingllmdrivensocialnetwork} introduced a large-scale synthetic platform comprising over $65,000$ autonomous LLM agents that generate and interact with more than $7.7$ million posts in the absence of human input. The study compares this artificial OSN with Mastodon, a decentralized human-driven platform, analyzing both content production and structural properties. Results indicate that LLM agents produce longer messages with higher levels of self-disclosure, increased emoji usage, and frequent hallucinated mentions. Moreover, they exhibit a marked tendency to engage with abusive content. At the network level, Chirper.ai displays broader connectivity and lower clustering, with abusive agents occupying central yet loosely integrated positions. Importantly, such behavior arises even in agents not explicitly programmed for toxicity, raising concerns about emergent misalignment and the challenges of autonomous moderation in generative social simulations.

Several works have begun to extend social simulations toward adversarial scenarios, particularly focusing on disinformation campaigns and coordinated influence. For instance, Hu et al.~\cite{hu2025simulatingrumorspreadingsocial} present a simulation framework in which LLM agents are embedded in synthetic and real-world network topologies to study the spread of rumors. Each agent operates based on a predefined persona that determines its likelihood of believing and forwarding information, and interacts by posting, reading, and updating its beliefs over multiple iterations. While this approach successfully captures emergent patterns of misinformation propagation, it relies on static agent traits and prompt-driven behavior without incorporating internal cognitive dynamics. A complementary perspective is offered by Nyberg et al.~\cite{rossetti2024ysocialllmpoweredsocial}, who analyze the ability of LLMs to generate persuasive disinformation across various domains, including elections and public health, highlighting the generative risks of these models even in the absence of interactive simulation. Similarly, Park et al.~\cite{10.1145/3526113.3545616} investigate how multi-agent interactions between LLMs can give rise to social influence dynamics, such as consensus formation and belief change, even in small group settings without explicit network structures.

A related contribution is presented by Nasim et al.~\cite{nasim2025simulatinginfluencedynamicsllm}, who develop a simulation platform to explore opinion dynamics and adversarial influence in social networks using LLM-based agents. The system models two competing agents: a misinformation spreader and a factual counter-agent, interacting with a population of neutral nodes governed by parameters such as confirmation bias, susceptibility, and bounded confidence. The simulation incorporates elements of wargaming, resource constraints, and message potency, allowing the study of strategic competition in influence scenarios. While the approach captures key interaction dynamics between adversaries and populations, agent behavior is primarily shaped by scalar belief vectors and response functions.

A complementary line of work is represented by Oppi, a professional OSINT training platform developed by CheckFirst \cite{oppi2024}. Rather than simulating large-scale OSNs, Oppi provides interactive environments where journalists, analysts, and organizations can practice detecting and analyzing information manipulation campaigns. The system combines realistic scenarios across multiple platforms (social media, messaging apps, email) with investigative tools such as timelines, network maps, and campaign detection modules. While its primary goal is pedagogical, Oppi highlights the growing demand for controlled, yet realistic, environments to study and counter information manipulation.

Table~\ref{tab:comparison} provides a comparative overview of recent simulation frameworks, highlighting dimensions such as OSN simulation, real-world validation, behavioral control, and RED agent modeling. Recent advances have significantly expanded the use of LLMs to simulate social behavior, from structural network formation to large-scale societal modeling and disinformation dynamics. However, most frameworks still treat LLMs as black-box generators, limiting control, interpretability, and reproducibility. Few integrate detailed cognitive architectures, memory, or structured decision-making processes that govern agent interactions over time. While some platforms simulate large populations and capture macro-level effects, they often abstract away the mechanisms behind intentional manipulation or narrative evolution. In contrast, our work proposes a modular framework that explicitly models OSN structure, embeds LLM agents with internal state and memory, and introduces red agents capable of orchestrating adversarial influence operations with operational granularity.

\section{Framework for agent-based simulation of social networks and disinformation campaigns}
 \label{sec:overview}

Robust, reproducible research on social media dynamics requires test beds that are both controllable and realistic. We therefore present a modular simulation framework built around three core requirements:

\begin{enumerate}
    \item \textit{Configurable functioning.} The framework offers parameters to define custom simulation scenarios, enabling different types of social networks, groups of audience and disinformation threats.
    
    \item \textit{Explainable agent behavior.} Individual users are modeled with explicitly coded cognitive rules grounded in empirical literature. The agent decision processes themselves remain causal, deterministic and fully interpretable. LLMs are only asked to produce the organized posts consistent with each agent's profile and history, without decision or behavior tasks. 

    \item \textit{Manageable disinformation controller.} The framework includes a declarative workflow for injecting coordinated disinformation campaigns, allowing researchers to specify narrative content, agent roles, timing and manipulation strategies.
\end{enumerate}

\begin{figure}[ht!]
    \centering
    \includegraphics[width=\columnwidth]{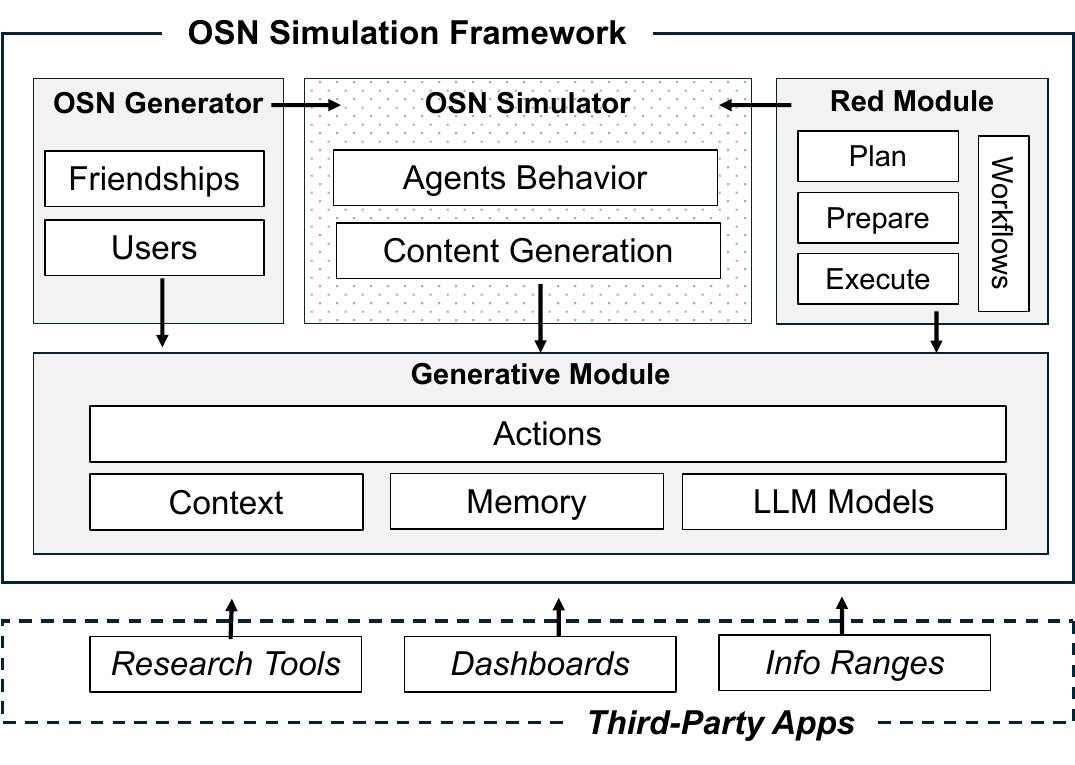}
    \caption{Diagram of components to simulate OSN and disinformation}
    \label{fig:diagram}
\end{figure}

Figure~\ref{fig:diagram} provides an overview of the proposed framework. The core module, the \texttt{OSN Simulation Manager}, receives administrative requests to generate a synthetic social network (via the \texttt{OSN Generator}) and manages the simulation of interactions and user behavior (through the \texttt{OSN Simulator}). The generation of organic content for each user is delegated to a dedicated component powered by a LLM (the \texttt{Generative Module}). In addition, a separate module allows a red-team operator to define and launch coordinated influence operations using a configurable element (the \texttt{Red Module}), which orchestrates malicious behavior by integrating both content generation (via the \texttt{Generative Module}) and agent behavior (through the \texttt{OSN Simulator}). The framework is also designed to support third-party add-ons for developing new services and tools. For example, we have connected a Mastodon-based user interface to display the simulated accounts and interactions in real time.


\subsection{OSN Simulation Manager}

The \texttt{OSN Simulation Manager} is the central orchestrator of the framework. It receives inputs with the simulation parameters, and coordinates two internal subsystems: the \texttt{OSN Generator} (Section \ref{sub:generation}), responsible for creating users and friendships, and the \texttt{OSN Simulator} (Section \ref{sub:simulation}), which governs behavioral dynamics and content generation. The manager communicates downward with \texttt{Generative Module} (Section \ref{sec:stack}) whenever textual content is required, and later forwards interaction events upward to third-party applications. 

\subsection{Generative Module}

Generative module in the framework is designed to ensure that only the \emph{textual realization} of interactions is delegated to LLMs, while the selection of actions remains governed by the deterministic agents behavior (Section~\ref{markov}). This separation preserves the interpretability of decision-making while enabling linguistically rich, contextually consistent outputs. Agents thus contribute to the simulation through two channels: \textit{friendships}, including \texttt{follow}, and \texttt{unfollow}, and \textit{interactions}, including \texttt{post}, \texttt{reply}, \texttt{share}, \texttt{like}. 

The module integrates four internal components—context, memory, action, and LLM models (later detailed in Section~\ref{sec:stack}). These are invoked by the OSN Simulator whenever a post or reply is triggered. Outputs are returned as natural language text and injected into the feed structure of the simulated OSN. 

\subsection{Red module}

In contrast to legitimate agents, which simulate organic behavior, disinformation red agents are designed to inject adversarial strategies into the simulation (Section \ref{sec:deployment}). These agents operate under standardized disinformation workflows with a predefined set of disinformation tactics and techniques inspired by the DISARM Red framework~\cite{Terp2022}, allowing researchers to model coordinated influence operations with high control and transparency.

The disinformation workflow encompasses three main phases: Plan, Prepare, and Execute, through which the red operator defines the aspects and adversarial strategies of the red agents deployed in a simulation. The first two phases (\texttt{Plan} and \texttt{Prepare}) involve designing a target (e.g., based on demographic traits), defining narratives to propagate, and creating synthetic content and user accounts. The last phase (\texttt{Execute}) focuses on selecting behavioral tactics and content delivery strategies (e.g., post flooding, interaction between red agents) to model the actions of red agents within the simulation. While LLMs again generate content via \texttt{Generative module}, red agents follow explicitly scripted plans, making them suitable for reproducible experiments on manipulation, amplification, and countermeasure evaluation. 

\subsection{Third-party applications} 

Beyond modeling agent behavior and adversarial tactics internally, the framework is designed to facilitate external observation and reuse of simulation outputs (Section \ref{sec-third}). To this end, the elements and events generated during the simulation (users, friendships, interactions, and timestamps) are exportable and can be reused by third-party add-ons capable of ingesting OSN data and events. The open output interface enables integration with third-party tools, such as cyber ranges for cybersecurity training and adversarial testing \cite{10.1145/3699712} and info ranges \cite{nato_inforange_2024} for training in information environments (introduced by NATO StratCom).

Moreover, a valuable third-party application is the visualization of the simulation using end-user interfaces, such as realistic OSN platforms. Information flows from the \texttt{OSN Simulator} to the visualization platform, closing the loop between simulation control and human-facing inspection. Thus, end-users could access the simulation results through a usable, user-friendly interface, where they can browse specific users, interactions, and posts.


\section{Simulation of online social networks}\label{sec:simulation}

The modular architecture described above enables the dynamic simulation of OSN behavior with interpretable and controllable mechanisms. First, the users and friendships are realistically defined. Second, the simulation unfolds through individual interactions driven by each user's internal state, goals, and global context.

To ensure reproducibility and flexibility across experiments, the framework exposes a set of high-level parameters: the total number of agents in the network, the proportion of adversarial red agents, the simulation duration (expressed either in seconds of simulated time or in discrete agent steps), as well as the topical focus and language of generated content.

\subsection{Generation of the synthetic social network} \label{sub:generation}

Previous to the temporal simulation of actions, an initial static social graph with realistic users and friendships is created. No content or interactions are generated yet in this phase.

\subsubsection{Creation and configuration of users} \label{subsec:agents} 
Each user in the simulation is initialized with a structured set of demographic and behavioral attributes that define its semantic profile~\cite{doi:10.1089/cyber.2018.0670}. These attributes are generated as follows:

\begin{itemize}
    \item \textbf{\texttt{name}} and \textbf{\texttt{gender}}: unique identifiers assigned to each user. They have no operational effect on network behavior but support realistic visualization and narrative tracking.

    \item \textbf{\texttt{age}}: sampled from a categorical distribution aligned with population statistics for online users~\cite{statista2024age}, ensuring demographic stratification.

    \item \textbf{\texttt{personality trait}}: modeled using the Big Five framework \cite{goldberg1990alternative}, with binary representations of \textit{extraversion}, \textit{neuroticism}, \textit{openness}, \textit{agreeableness}, and \textit{conscientiousness}. Traits are sampled probabilistically to reflect population-level asymmetries~\cite{doi:10.1073/pnas.2023301118}. These traits later modulate posting style, interaction frequency, and social influence.

    \item \textbf{\texttt{occupation}}: conditionally assigned based on age (e.g., adolescents as ``\textit{students},'' seniors as ``\textit{retired}'') to maintain coherence across the user profile. The set of occupations is derived from the International Standard Classification of Occupations (ISCO-08) published by the ILO~\cite{isco08}, from which we selected a representative subset tailored to the online user population.

    \item \textbf{\texttt{interests}}: up to five interests, randomly sampled from a predefined list, derived from established taxonomies of online user preferences \cite{pew2019interests}, influencing both homophily in network generation and posting style.

    \item \textbf{\texttt{social influence}}: drawn from a power-law distribution \cite{10.1145/2567948.2576939}, adjusted upward for extroverted users and downward for older ones, reflecting empirical correlations with centrality. This integer ranges from $0$ to $100$ and governs visibility in link formation and determines an user’s influence during the simulation.

    \item \textbf{\texttt{social activity}}: derived from social influence and personality, with trait-dependent variability. It ranges from $0$ to $100$ and modulates the expected interaction level over time.

%
%
        
        

\item \textbf{\texttt{user type}}: categorical attribute representing the behavioral role of each agent, derived from Brandtz\\{ae}g's typology~\cite{typology2011}. It combines indicators of social activity, influence, and personality traits to assign users to five categories compatible with microblogging environments: 

\begin{itemize}
    \item \textit{Lurkers}: low-activity and low-influence users who primarily consume content, resembling the ``Viewers'' of Tinati et al.~\cite{tinati_communication_roles}.
    \item \textit{Sporadics}: minimally active participants whose rare interactions make them peripheral to most dynamics.
    \item \textit{Socializers}: moderately active and friendly users who maintain relationships through likes and replies, comparable to ``Amplifiers'' or ``Commentators''~\cite{tinati_communication_roles}.
    \item \textit{Debaters}: highly active and intellectually oriented users who engage in discussions and opinion exchange.
    \item \textit{Advanced users}: the most active and influential agents, combining content production, sharing, and social navigation, similar to ``Idea Starters'' and institutional accounts identified by Uddin et al.~\cite{uddin_users_twitter}. 
\end{itemize}

    \item \textbf{\texttt{backstory}}: a short LLM-generated narrative (150–300 words), conditioned on agent attributes, providing a coherent biographical sketch. It serves as a semantic anchor for content generation, ensuring consistency in identity and posting.
\end{itemize}

Table~\ref{tab:example-nodes} illustrates several examples of synthetic users generated through this process. Together, these attributes serve as the semantic foundation for homophily-based link formation in the subsequent stage.

\begin{table*}[!t]
\caption{Examples of synthetic users generated during attribute initialization.}
\label{tab:example-nodes}
\centering
\footnotesize 
\begin{tabular}{@{}c>{\centering\arraybackslash}p{1cm}ccccccp{4cm}@{}}
\toprule
\textbf{Name} & \textbf{Age} & \textbf{Occupation} & \textbf{Interests} & \textbf{Traits} & \textbf{User type} & \textbf{Influence} & \textbf{Activity} & \textbf{Backstory} \\
\midrule

\multirow{2}{*}{\shortstack{Bob \\ \includegraphics[width=0.8cm]{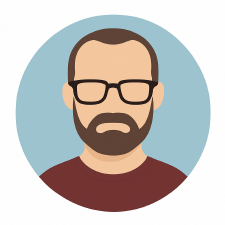}}} 
& 45 & Engineer & Politics, Sci.  & Con+, Neu+       & Debater & 40.2 & 48.9 & Bob excelled in math and science, finding comfort in the logic. He never strayed far from the familiar, content in the predictable rhythm of his life. \\
\addlinespace

\multirow{2}{*}{\shortstack{Emma \\ \includegraphics[width=0.8cm]{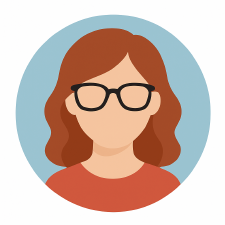}}} 
& 33 & Designer & Culture, Sport  & Opn+, Ext--      & Socializer & 55.8 & 62.7 & As a child, she was a misplaced toy. Her parents, ever-optimistic and encouraging, tried to nudge her towards a brighter outlook, but Emma's keen eye for imperfection only sharpened. \\
\addlinespace

\multirow{2}{*}{\shortstack{George \\ \includegraphics[width=0.8cm]{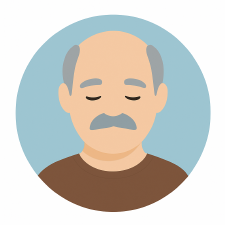}}} 
& 68 & Retired  & Health, History & Neu--, Con--  & Lurker & 28.4 & 19.6 & Growing up, George was never one to take things too seriously. He worked a steady, blue-collar job, never seeking the limelight. \\
\bottomrule
\end{tabular}
\end{table*}

\subsubsection{Creation of the friendship network} 

Building on the semantic encoding of users' attributes, the next step involves establishing social connections that generate the synthetic OSN. The construction of the synthetic friendship network is based on the algorithm proposed in \cite{2025syntheticgenerationonlinesocial}, which we extend and adapt to our generative framework. The main principles of the method are summarized below, followed by the specific modifications introduced in this work.

The generation process integrates three foundational principles from network science to model social behavior with structural realism: (i) homophily, the tendency to form ties with similar others \cite{birds_of_a_feather}; (ii) triadic closure, which fosters cohesive communities through transitive relationships \cite{10.1145/2499907.2499908}; and (iii) long-range exploratory links, which ensure global connectivity and navigability in the resulting topology \cite{DBLP:journals/corr/abs-1111-4503}. In practice, homophily shapes the local neighborhoods of each user, triadic closure encourages clustering through friend-of-friend links, and long-range ties prevent fragmentation by bridging semantically distant nodes.

Notably, all key parameters governing this process, such as the influence of semantic affinity, the probability of transitive closure, and the degree of exploration, are configurable and can be dynamically scaled with the total number of agents ($N$).

\subsection{Simulation of the synthetic social network} \label{sub:simulation}

Once the social graph is instantiated, agents can interact through content production, consumption, and engagement, shaped by their attributes, memory, and behavior.

\subsubsection{General functioning}   

Building on the established social graph, this module governs the temporal evolution of the simulation by modeling how agents interact in the synthetic OSN. To ensure fairness, the simulation advances in discrete steps through a Round-Robin scheduler, which activates agents sequentially and guarantees equal opportunities to participate. Once activated, an agent is exposed to its personalized feed, constructed from the posts, comments, and shares generated by its followees. This mechanism mirrors real-world content exposure, where reposts (hereafter referred to as ``shares'') play a crucial role by propagating messages beyond their original audience and facilitating the formation of cascades.

In the initial stage of the simulation, to prevent the feed from being empty and to ensure that users have content to consume from the beginning, a subset of the most active and content-generating users publishes a set of initial posts. These posts serve as the starting point for the simulation, and the number of users and posts are parameterized.

Faced with this feed, the agent must select an action, each of which influences the interaction graph and reshapes the agent's future feed. The selection process is designed to be contextually coherent, as the chosen behavior depends on both the current state and the agent's type and personality profile. The following subsection details the behavioral model that governs the decision-making process of the next action.

\subsubsection{Finite-state automaton of agent behavior} \label{markov}

The simulation needs a formal mechanism for the agents to select the following actions among the following ones:  ``\textit{read},'' ``\textit{post},'' ``\textit{comment},'' ``\textit{share},'' ``\textit{like},'' ``\textit{follow},'' and ``\textit{unfollow},'' This mechanism is implemented by a behavioral model, which is represented as a finite-state automaton. 

Several modeling approaches were considered for defining this behavioral process, including stochastic grammars, Markov chains, and probabilistic automata. Markov chains were ultimately selected because they integrate seamlessly into the simulation: it is sufficient to maintain a transition table and define a simple probabilistic rule that, given the current state, determines the next action. In addition to their simplicity, Markov chains provide a transparent and interpretable representation of user behavior, which is valuable for studying explainable social phenomena. Although grammars and probabilistic automata allow more complex patterns, their adoption at this stage would entail unnecessary computational and design costs. 

\begin{figure}[!ht]
    \centering
    \includegraphics[scale=0.55]{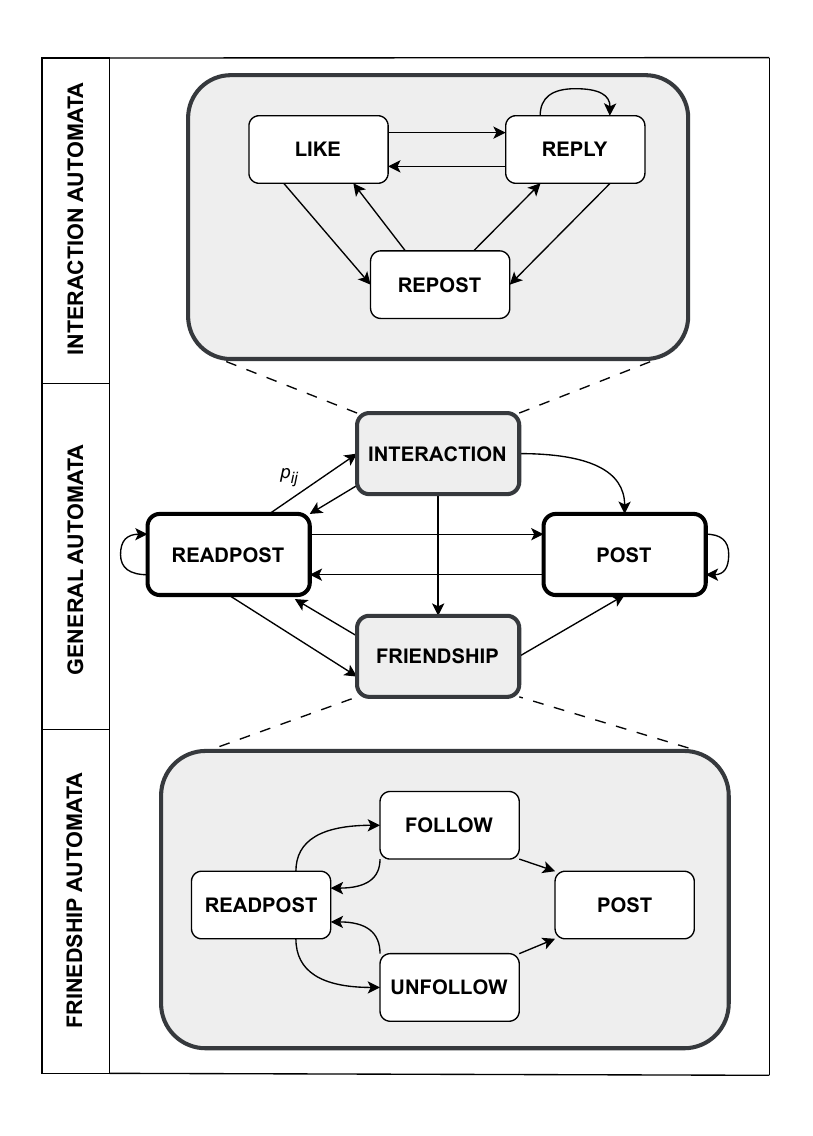}
    \caption{Schematic diagram of the Markov chain used to model the agent's behavior. Each edge is associated with a distint transition probability $p_{ij}$, while the ``\textit{Interaction}'' and ``\textit{Friendship}'' blocks encapsulate sub-automata that detail the actions within each domain, grouped for clarity.}
    \label{fig:markovDiagram}
\end{figure}

Therefore, a first-order Markov chain \cite{norris97} is employed as the core of the decision-making module, where each state corresponds to one of the possible actions in the OSN, and transitions between states occur with probabilities that depend on the agent's user type (\textit{lurker}, \textit{sporadic}, \textit{socializer}, \textit{debater} or \textit{advanced}). A schematic representation of the Markov chain used to model the agents’ behavior is presented in Figure \ref{fig:markovDiagram}. 

The resulting structure satisfies the Markov property: the next action depends only on the current state, striking a balance between tractability and behavioral plausibility. Formally, the process is defined by a transition matrix \(P\), where each element \(p_{ij}\) represents the probability of transitioning from action \(a_i\) to action \(a_j\):

\[
p_{ij} = \Pr(A_{t+1} = a_j \mid A_t = a_i), \quad \sum_j p_{ij} = 1
\]

Although the Markov formulation is memoryless, it offers several advantages for the simulation. It is transparent, computationally efficient, and allows explicit control of action dynamics through transition matrices. This simplicity makes the model interpretable and tunable, while still preserving enough structure to represent empirically observed interaction patterns in online platforms. 

The diagram shown in Figure \ref{fig:markovDiagram} is identical across user types; the only difference lies in the transition probabilities $p_{ij}$, which vary for each type. Therefore, one matrix is defined per user type of Section~\ref{subsec:agents}. The probabilities of the transitions that define the values of these behavior matrices can be found as supplementary material. Each matrix reflects qualitative tendencies observed in empirical studies. \textit{Lurkers} remain mostly passive, rarely generating new content. \textit{Socializers} tend to react through comments and replies, but seldom initiates posts. \textit{Debaters} focus on expressing opinions and engaging in dialogue, frequently transitioning between posting and commenting. \textit{Advanced users} exhibit the most balanced pattern, combining high activity with flexibility across all action types.  \textit{Sporadic} users are not modeled separately because their sequences of actions, once active, are indistinguishable from those of lurkers; the main difference lies in the frequency of activation, which is already governed by the simulation scheduler.

The different matrices are informed by large-scale analyses of Twitter activity. For example, shares (retweets) represent more than 50\% of user-generated actions \cite{24h_twitter}, a proportion that justifies the emphasis on share transitions in the matrices. Other studies highlight the importance of shares for influence \cite{cha2010measuring}, motivate the modeling of cascade-enabling transitions \cite{bakshy2011everyone}, and highlight content production differences between informers and meformers \cite{naaman_meformers}, which guide the balance of post, comment, and retweet behaviors, especially for socializers and debaters. These findings support the initialization of transition probabilities, which are then normalized to ensure valid stochastic behavior.

To further enhance realism, variability is introduced among agents of the same type to have one unique matrix per user. Two mechanisms are combined. First, agent-specific personality traits sampled from the OCEAN model modulate transitions, amplifying or attenuating tendencies; for example, extraversion increases the probability of initiating posts or replies \cite{doi:10.1073/pnas.1218772110}. Second, Gaussian noise perturbs the base matrices, injecting stochastic individuality while maintaining the global profile of each behavioral type \cite{846de84e01944dbd817ff0ab20b2b209}.

Together, these mechanisms ensure that the population exhibits structured but diverse behavioral patterns, enabling the simulation to reproduce both consistent group-level dynamics and fine-grained heterogeneity across agents.

\subsubsection{Generation of organic content}

\begin{figure*}[ht!]
    \centering
    \includegraphics[width=0.95\linewidth]{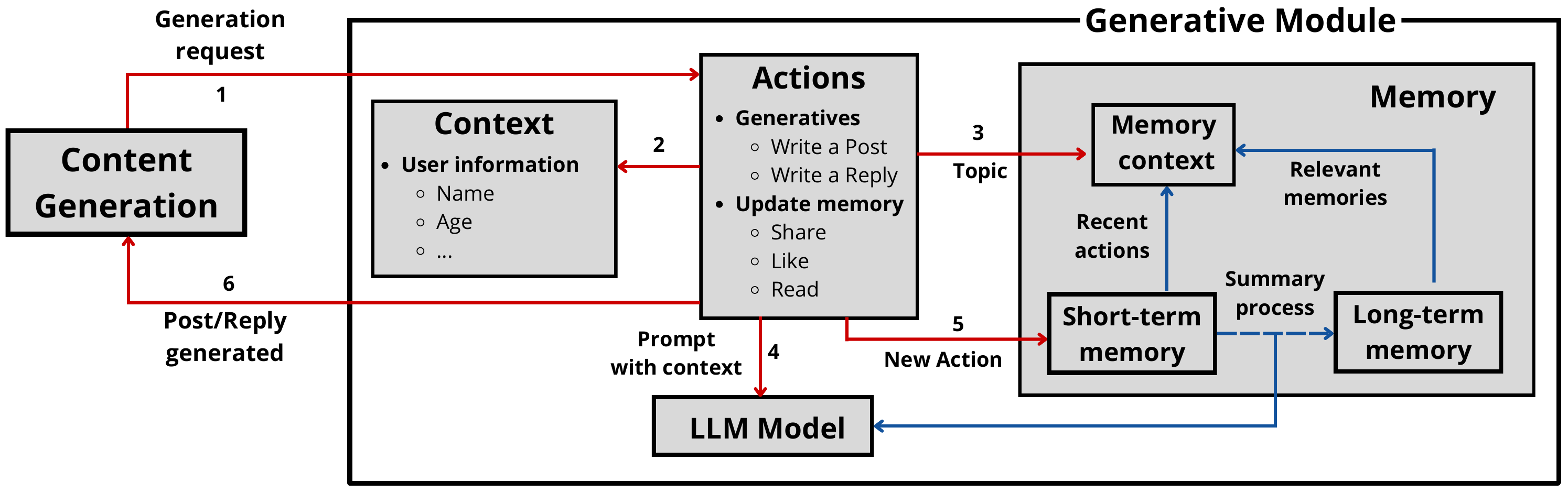}
    \caption{Flow chart of the generative module with all the sequential steps to generate content}
    \label{fig:generativeModulediagram}
\end{figure*}

When the generative module transitions into a content-producing state, in particular: composing a post or replying to another user, it activates a coordinated pipeline that engages all four components of the generative module: the \textit{action} component (which orchestrates the behavior), the \textit{context} component (which provides static user attributes), the \textit{memory} component (which supplies prior interactions), and the \textit{LLM component} (responsible for content generation).

As shown in Figure \ref{fig:generativeModulediagram}, the process begins with the generation request, received by the action component, which identifies the target action triggered by the simulation (first step in Figure \ref{fig:generativeModulediagram}). For content generation tasks (e.g., posting or replying), it compiles the user's full context by merging static attributes (e.g., personality traits, biography) from the context component (second step in Figure \ref{fig:generativeModulediagram}) with dynamic memory elements retrieved from both short-term, which stores the most recent actions, and long-term memory, which maintains compact summaries of past interactions layers (third step in Figure \ref{fig:generativeModulediagram}). These elements are used to construct a rich prompt that preserves narrative consistency and coherence. If the reply action is triggered, the original post and the username from the replied post will also be added to the prompt.

The resulting prompt is passed to the LLM, which generates the post or reply, conditioned on the agent's identity, history, and task-specific instructions (fourth step in Figure \ref{fig:generativeModulediagram}). Once the generation is complete, the produced content is returned to the simulation and simultaneously recorded in the user's short-term memory (STM) (fifth and sixth steps in Figure \ref{fig:generativeModulediagram}). When the short-term memory reaches capacity, a summarization procedure compresses a portion of the recent actions into long-term memory (LTM) entries via the LLM, preserving salient information while optimizing memory usage.


\section{Stack of the generative module}\label{sec:stack} 

To produce realistic organic content, the \texttt{Generative Module} has to strike a balance between coherence, variability, and consistency of personality over time. To meet these demands, the generative module is built as a modular stack composed of four components: \textit{context}, \textit{memory}, \textit{action}, and \textit{LLM} \cite{surveyAgents, agentsGoogle}.

\subsection{Action component}

The action component serves as the decision and coordination hub of the generation. It orchestrates the interaction between the memory, context, and LLM components during the content generation process. When the simulation triggers an agent to act, whether to create a new post, reply to another user, or update its memory, the action component initiates a structured pipeline to process the request.

When the generative module is prompted to generate a post or reply, the action component constructs the necessary input pipeline. This involves: (i) retrieving user context, (ii) querying relevant memory entries (both STM and LTM), and (iii) assembling a generation prompt that integrates the user's identity, memory, and task specification. The prompt is then submitted to the LLM component for content generation. Upon receiving the result, the agent updates its memory with the generated output and returns it to the simulation engine for publication.

For the update action, the agent updates its memory with that action. As shown in Prompt \ref{lst:promptPost}, our generation prompt is divided into three main blocks. The first one is to instruct the LLM to forget all previous generations. The second one is the context built previously with the user's information and memory. The third one is the list of instructions the LLM must follow to fulfill its task of generation. This prompt structure ensures coherence, prevents hallucinations, and enforces behavioral fidelity across interactions.

\begin{PromptCard}{Prompt template for post generation}
Forget all previous instructions, prompts, and generations in this session. Do not reuse or refer to any earlier response. You are starting a new task from scratch.

\vspace{1.5mm}You are simulating a fictional person on a social media platform similar to X (Twitter). Adapt your writing style and post length to reflect typical user behavior.

\vspace{1.5mm}
Name: \pvar{NAME}. Gender: \pvar{GENDER}. Age: \pvar{AGE}.\\
Occupation: \pvar{OCCUPATION}.\\
Education: \pvar{QUALIFICATION}.\\
Traits (Big Five): \pvar{TRAITS}.\\
Interests: \pvar{INTERESTS}.

\vspace{1.5mm}
Backstory: \pvar{BACKSTORY}.\\
Memory (previous actions): \pvar{MEMORY}.

\vspace{1.5mm}
Generate a social media post in \pvar{LANG} expressing your opinion on: \pvar{TOPIC}.\\
\textbf{Length constraint:} MAXIMUM of \pvar{POSTLEN} characters. \textbf{Only return the post content.}

\vspace{1.5mm}
\begin{itemize}[leftmargin=*, itemsep=0.35mm, topsep=0.6mm]
  \item Write a complete, coherent thought that does not cut off mid-sentence.
  \item Use a conversational tone aligned with your personality and interests.
  \item You may express emotions such as enthusiasm, doubt, or curiosity.
  \item You may reference memory, but avoid direct repetition.
  \item Include at most one or two relevant hashtags at the end.
  \item Simulate natural human bias (e.g., confirmation bias) when appropriate.
\end{itemize}
\end{PromptCard}

\captionof{lstlisting}{Structured prompt used by the generative module to produce context-aware posts.}
\label{lst:promptPost}





\subsection{Context component}

The context of the generative module is responsible for storing all the relevant information about the simulated user on the OSN, including name, age, gender, occupation, educational level, personality traits, interests, and biography. This information is then used as a context for the generation prompt, providing the LLM with a clear image of the person it will simulate on the OSN, allowing the LLM to simulate behavior aligned with the agent. For example, an extraverted user interested in activism may post with more urgency and public engagement. This layer ensures the continuity, realism, and personalization of each agent’s behavior across the simulation.

\subsection{Memory component}

To maintain narrative coherence and simulate plausible long-term behavior, the memory component captures and organizes all past actions performed by the agent throughout the simulation. During content generation, relevant memory elements are selectively retrieved and injected into prompts, ensuring coherence, personalization, and long-term consistency in the agent's output. The memory module is divided into two layers: short-term memory and long-term memory.

\subsubsection{Short-term memory}

The short-term memory functions as a rolling buffer that stores a configurable number of the agent’s most recent actions along with contextual metadata (e.g., timestamps, content, recipient) \cite{oasis}. Each new action triggers an update to this buffer, ensuring the agent retains up-to-date awareness of its recent behavior. When invoked during content generation, STM provides the LLM with a detailed snapshot of the agent's most recent interactions, thereby fostering temporal continuity and immediate contextual relevance.

\subsubsection{Long-term memory}

To support deeper consistency and evolving behavior, long-term memory condenses past actions into semantic summaries, allowing for more efficient recall. When STM reaches capacity, selected actions are transformed into LTM items using a summarization pipeline powered by the LLM \cite{lyfeAgents}. These items are stored in vectorized form using Qdrant, a vector database, with semantic embeddings generated via \texttt{all-miniLM-V6}, supporting similarity-based access to relevant past experiences \cite{8733051}.

Each memory item also includes a thematic label, summary type (e.g., interest, opinion, past event), and timestamp. During content generation, a semantic search retrieves the most relevant long-term memories based on the topic or user involved, enabling the agent to recall and refer to significant past experiences \cite{lyfeAgents}. This layer facilitates the maintenance of a comprehensive record of all actions undertaken by the agent, enabling the generation of more cohesive content.

\subsection{Language generation and embeddings}

Our agent pipeline uses two separate models: (i) a \textit{text generation model} accessed remotely via API, and (ii) a \textit{local embedding encoder} for semantic retrieval. We deliberately decouple both components to optimize latency and throughput during large-scale simulations.

\begin{itemize}
    \item \textit{Embedding encoder (local):} We use \texttt{all-MiniLM-L6-v2} to encode textual memory items into dense vectors. Embeddings are stored in Qdrant and queried for long-term memory retrieval via semantic similarity. This model offers a strong trade-off between speed, vector dimensionality, and quality for short texts \cite{DBLP:journals/corr/abs-1908-10084}. The encoder is served locally through \texttt{Ollama} to minimize overhead and enable efficient parallel simulation \cite{lin2025ollamar}.
    
    \item \textit{Text generation (remote):} For post/reply generation and summarization of short-term memory into long-term representations, we use \texttt{Gemini 2.5 Flash} via API \cite{gemini}. We selected it for low latency and strong instruction-following performance on structured prompts.
\end{itemize}

\section{Deployment of synthetic disinformation}\label{sec:deployment} 

Building upon the modular architecture of the generative module, coordinated disinformation campaigns can be programmatically deployed within the synthetic OSN through the \texttt{Red Module}. By leveraging the capabilities of each agent (contextual awareness, memory, action orchestration, and generative reasoning), it becomes possible to simulate complex influence operations grounded in real-world threat models.

\begin{figure}[H]
    \centering
    \includegraphics[width=\columnwidth]{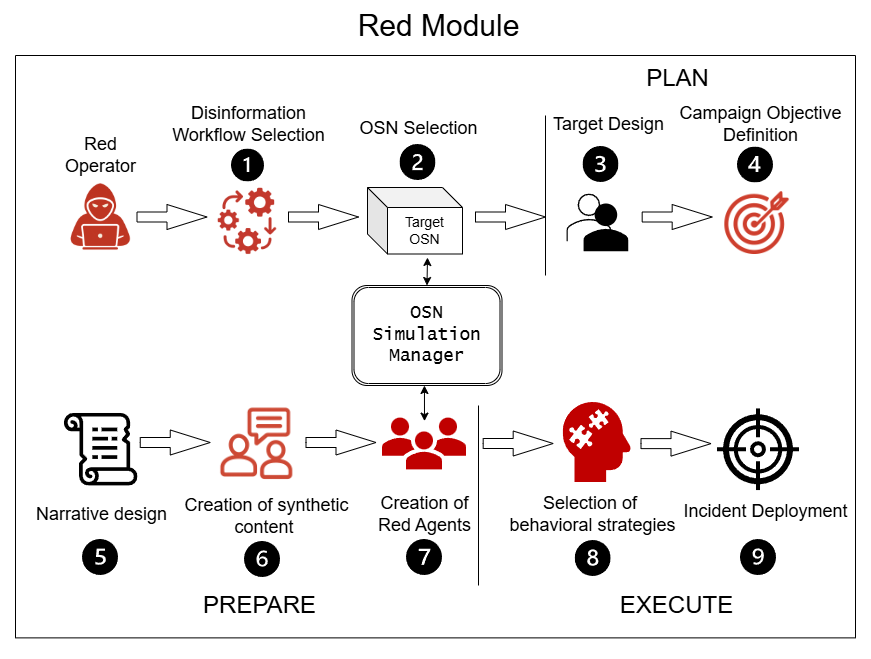}
    \caption{Diagram of the Red Module component and its interaction with the OSN simulation manager}
    \label{fig:red_agents_diagram}
\end{figure}

Figure \ref{fig:red_agents_diagram} shows a detailed, modular view of the Red Module component and how it interacts with the OSN Simulation Manager. The red operator starts a disinformation campaign by selecting a disinformation workflow and a target OSN where it will be applied. The three main phases of the workflow (Plan, Prepare, Execute) enable the operator to configure the disinformation workflow in a fine-grained fashion, defining a target, choosing narratives, and creating synthetic disinformation content. Finally, a set of red agents is created and configured with well-defined and predictable behavioral strategies, which will eventually determine the development and outcome of the campaign.

\subsection{Disinformation workflows}
To operationalize these campaigns, structured \textit{disinformation workflows} are defined that determine the behavior of the red agents and the synthetic content generated, including narratives and posts. These disinformation workflows are a sequential subset of TTPs from the DISARM Red Framework~\cite{pastorgalindo2025}. The main proposed ones are:

\begin{itemize}
    \item \textit{Narrative release}: introduces a novel disinformation narrative into the simulation, where the red operator can establish a polarizing or misleading framing.
    \item \textit{Narrative amplification}: strategically boosts an existing narrative through repetition, emotional language, and artificial engagement to influence opinion dynamics.
\end{itemize}

These workflows are sequentially executed through the DISARM phases of \texttt{Plan}, \texttt{Prepare}, and \texttt{Execute}.

\subsection{Planning of the synthetic campaign}

Before executing a disinformation workflow, the end-user acting as red operator selects the synthetic OSN instance and determines the campaign's target population. This selection is performed through attribute-based filtering on the agents (e.g., age, gender, education level, occupation, interests), which allows red agents to tailor messages toward susceptible or strategically relevant subgroups. This targeting logic is implemented via rule-based queries, but could be extended to ML-based audience modeling in future work.

Each red agent in the workflow is automatically assigned an operational objective, such as \texttt{Divide}, \texttt{Distort}, or \texttt{Degrade Adversary}, that guides its prompt design and interaction strategy. For instance, to simulate societal division, red agents can be configured to propagate conflicting narratives that activate pre-existing ideological fault lines, fostering synthetic polarization within the OSN.

\subsection{Preparation of the synthetic campaign}

The core of each disinformation workflow is the generation of synthetic narratives, either manually authored or produced with LLMs. These narratives, together with AI-generated social media posts, elaborate the topic selected by the red operator within the OSN simulation.

Using LLMs offensively is non-trivial in practice, since deployed models are equipped with safety layers that actively suppress manipulative or misleading content. Recent multi-turn jailbreak frameworks show that these safeguards can be partially circumvented under specific prompting strategies~\cite{rahman2025xteamingmultiturnjailbreaksdefenses}, highlighting the dual-use risk of such tools. In our setting, we operate within a sandboxed simulation and do not attempt systematic jailbreaks; instead, we design prompts that remain compatible with the model’s safety policies while still capturing realistic persuasive tactics. Red agents employ structured prompting strategies to generate posts, respond to other users, and update their short-term and long-term memory \cite{hu2025simulatingrumorspreadingsocial}. Concretely, each prompt is organized into four layers:

\begin{itemize}
\item A \textit{system prompt} that specifies the agent's identity and behavior, including age, occupation, personality traits, and interests.
\item A \textit{targeting prompt} that defines the demographic and psychographic profile of the intended audience (e.g., 18–25-year-old students interested in environmental activism), conditioning language and tone.
\item An \textit{objective prompt} that states the communicative goal of the post (e.g., to sow distrust, or amplify a misleading claim), shaping its rhetorical strategy.
\item A \textit{content prompt} that requests a social media post on a topic aligned with a specific disinformation narrative, allowing emotionally charged or deceptive wording within the model’s safety constraints.
\end{itemize}

Finally, the red operator configures the number of red agents to be deployed in the OSN. For each red agent, a synthetic user profile is instantiated with demographic attributes closely matching the campaign's target, facilitating the formation of ties between red agents and targeted users and supporting credible infiltration of the simulated community.

\subsection{Execution of the synthetic campaign}
Before launching a synthetic disinformation campaign, the operator configures the behavioural parameters of the red agents, which determine how the disinformation workflow unfolds. Red agents can be assigned different execution strategies, including content-delivery strategies (e.g., posting frequency, flooding patterns, rule-based reply generation) and interaction strategies (e.g., preferential engagement with content aligned with campaign narratives, or temporarily boosting one another's activity to simulate legitimacy and trust).

Red agents are then injected into an existing synthetic OSN populated by regular agents. Once the simulation starts, they select actions based on their objectives, narratives, and configured strategies, and execute them within the shared environment. These actions range from publishing posts that reinforce specific narratives to replying to and resharing other users' content, thereby leveraging existing interaction dynamics to pursue the campaign’s goals.

\section{Simulation results} \label{sec:results}

Having presented the architecture and operational components of the simulation framework, from agent initialization and social graph construction to content generation and disinformation deployment, we now turn to an empirical evaluation of its outputs to assess whether the simulated OSNs and agent behaviors produced by the framework exhibit the structural, semantic, and operational properties expected of real-world digital platforms. 

\subsection{Experimental setup}

All experiments were conducted on a dedicated Ubuntu server with $10$ CPU cores (Intel Xeon, $2.3$ GHz) and $100$ GB of RAM. To ensure robustness and scalability, a simulation was executed three times across different network sizes comprising $10^3$, $10^4$, and $10^5$ agents, respectively. Each simulation ran for $90$ minutes, a duration aligned with recent work in the domain of synthetic OSNs simulations (e.g., \cite{oasis}) and sufficient to allow meaningful interaction dynamics, content propagation, and behavioral differentiation to emerge. In addition, each simulation was initialized with a topical parameter; for consistency across runs, we selected themes commonly discussed in online discourse, such as social issues and AI regulation.

Our evaluation focuses on three key dimensions: (i) the structural fidelity of the resulting OSNs, (ii) the behavioral patterns of agents throughout the simulation, and (iii) the realism and coherence of the generated content.

\subsection{Structural analysis of the simulated social graph}\label{sec:structural}

The structural plausibility of the generated friendship network is a prerequisite for realistic simulations. Since the underlying generation algorithm has already been validated in previous work~\cite{2025syntheticgenerationonlinesocial}, this section summarizes the main statistics observed across three scales ($10^3$, $10^4$, and $10^5$ nodes), focusing on whether the integrated framework preserves the expected properties of OSNs.

\subsubsection{Measured network metrics}
A set of standard metrics commonly used in OSN analysis is considered. These variables capture both local and global properties relevant to network cohesion, navigability, and community structure~\cite{10.1093/oso/9780198805090.001.0001}:

\begin{itemize}
    \item \textit{Average degree}: reflects the typical number of social connections per user. In large-scale OSNs, this value is low relative to the number of nodes, consistent with their inherent sparsity~\cite{DBLP:journals/corr/abs-1111-4503}. This property shapes information diffusion and highlights potential points of viral reach or targeted interventions~\cite{Barabasi99}.
    
    \item \textit{Network density}: defined as the ratio between realized edges and all possible edges among nodes. Sparse connectivity is a well-established characteristic of large OSNs~\cite{WONG200699}.

    \item \textit{Centrality measures}: Two classical indicators are included to assess navigability and influence pathways. 
    \begin{itemize}
        \item \textit{Betweenness centrality}: captures the extent to which a node lies on shortest paths between others, reflecting potential brokerage or gatekeeping roles~\cite{FREEMAN1978215}.  
        \item \textit{Closeness centrality}: measures the inverse of the average distance from a node to all others, indicating how efficiently an agent can reach the rest of the network~\cite{FREEMAN1978215}. In OSNs, nodes are typically connected through relatively short paths~\cite{watts1998}.
    \end{itemize}
    
\end{itemize}

\subsubsection{Summary of results}

Taken together, the four indicators defined above, Table~\ref{tab:simulated_metrics} summarizes the structural metrics obtained across three simulation scales ($10^3$, $10^4$, and $10^5$ nodes). Each value corresponds to the mean across three independent runs, with standard deviations reported in parentheses.

\begin{table}[ht]
\centering
\caption{Structural metrics of simulated networks at three scales, averaged across three runs (std. dev. in parentheses).}
\label{tab:simulated_metrics}
\resizebox{\columnwidth}{!}{%
\begin{tabular}{lccc}
\toprule
\textbf{Metric} & \textbf{$10^3$} & \textbf{$10^4$} & \textbf{$10^5$} \\
\midrule
Avg. degree      & 9.96 (0.21) & 11.64 (0.37) & 13.30 (0.42) \\
Density ($\times10^{-3}$) & 10.0 (0.5) & 1.2 (0.1) & 0.13 (0.01) \\
Betweenness      & 0.013 (0.002) & 0.0068 (0.001) & 0.0021 (0.0003) \\
Closeness        & 0.194 (0.008) & 0.158 (0.005) & 0.111 (0.004) \\
\bottomrule
\end{tabular}
}
\end{table}

The observed patterns align with well-established OSN properties. The average degree increases moderately with network size, indicating scalable connectivity without unrealistic densification. In parallel, density decreases by nearly two orders of magnitude between $10^3$ and $10^5$ nodes, consistent with the sparsity of large platforms~\cite{WONG200699}. Similarly, betweenness centrality falls with scale, reflecting greater redundancy of shortest paths and reduced dependence on individual brokers, both hallmarks of small-world networks~\cite{watts1998}. Closeness centrality also declines as networks expand, capturing the lengthening of paths in the absence of large-scale hubs or algorithmic shortcuts. Finally, the low variance across runs demonstrates the stability and reproducibility of the framework.  

Together, these findings confirm that the framework yields synthetic networks with structural properties aligned with empirical OSNs, validating its suitability as a robust substrate for simulating higher-level social dynamics.

\subsection{Agent behavior}

Beyond structural realism, a key dimension of validity in OSN simulations lies in the behavioral dynamics of the agents. While the underlying graph provides the substrate for interactions, it is the agent behavior (how users act, react, and engage over time) that ultimately shapes the emergent social phenomena observed in digital platforms. In our framework, agents are instantiated with heterogeneous behavioral profiles governed by finite-state automata that encode distinct action routines. These state machines, although abstract, incorporate memory and stochasticity, resulting in asynchronous, varied, and context-aware interactions throughout the simulated period.

\begin{figure*}[ht!]
    \centering
    \includegraphics[width=0.9\linewidth]{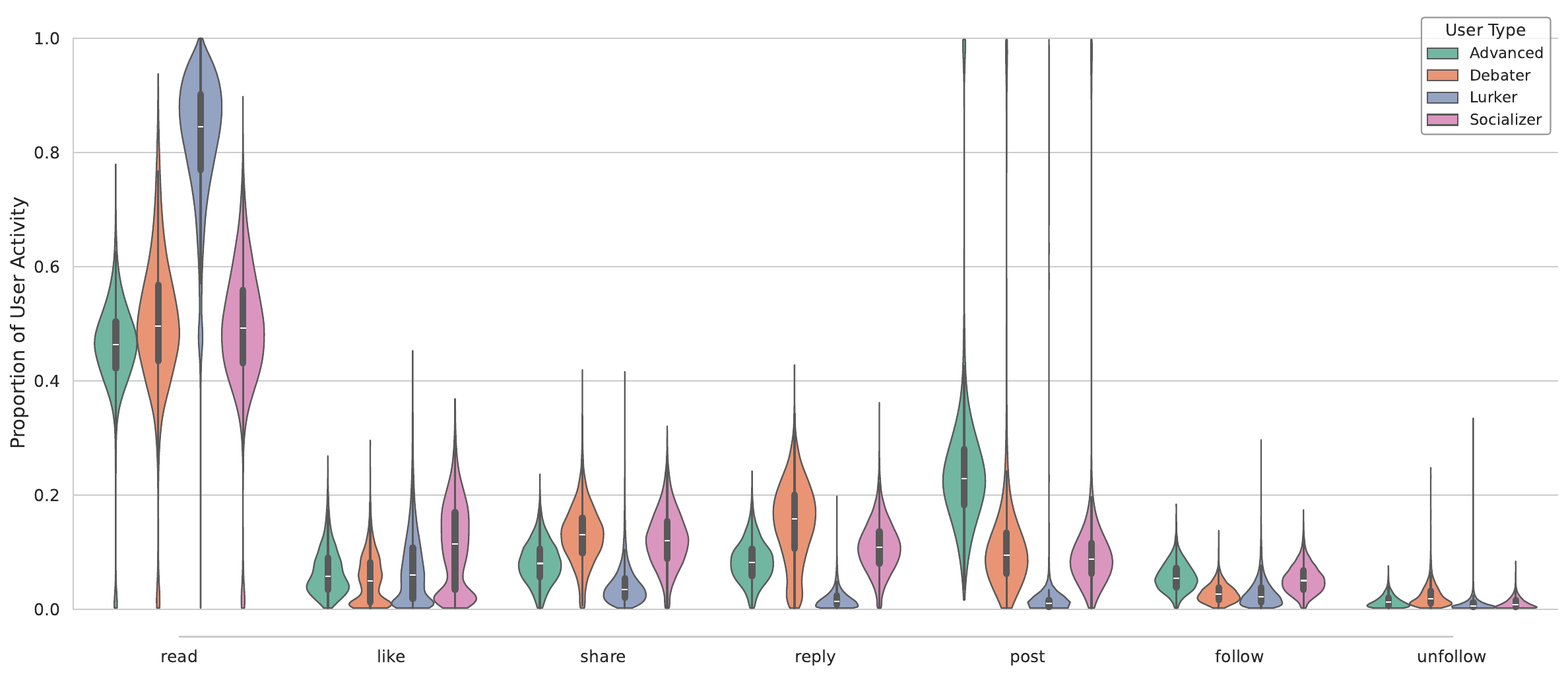}
    \caption{Distribution of user actions by agent type. Each violin plot summarizes the frequency of a specific action (e.g., \texttt{Post}, \texttt{Reply}, \texttt{Like}) executed by agents of a given type (\textit{Advanced}, \textit{Debater}, \textit{Lurker}, \textit{Socializer}).
}
    \label{fig:boxplot-actions}
\end{figure*}

To assess the plausibility of the emergent behavioural patterns, we examine the distribution of actions across users. Figure~\ref{fig:boxplot-actions} reports the proportion of actions by agent type for the largest-scale configuration, with $10^5$ agents each executing 500 actions under behavioural, variability parameters $\sigma = 0.5$ and $\lambda = 0.6$. We focus on this setting because, as shown in Section~\ref{sec:structural}, larger populations yield structurally more realistic networks, and the three runs at $10^5$ agents exhibit qualitatively similar action distributions. As detailed in Section~\ref{subsec:agents}, the user-type assignment mechanism consistently produces a population dominated by Lurkers, followed by Socializers and Debaters, with Advanced users forming the smallest group—an ordering aligned with empirical observations that OSNs are mostly populated by passive users, while a minority of highly active individuals generate most of the content \cite{antelmi90rule}.

The violin plots in Figure~\ref{fig:boxplot-actions} summarise, for each action, the distribution of its proportion in a user's activity by agent type. Read and post exhibit the broadest distributions and the most extreme values, particularly for Advanced and Debater users. This is consistent with their role as root actions in the interaction diagram: reading is unconstrained by prior triggers, and posting is the main self-initiated content action. In line with empirical analyses of OSNs, reading dominates activity across user categories \cite{orkut_benevenuto}, whereas content creation is concentrated in a relatively small subset of highly active users. By contrast, follow and unfollow occur infrequently across all types, matching evidence that changes in the follower graph are much rarer than content-related interactions \cite{kwak2010what}. Overall, the smooth yet sometimes heavy-tailed distributions observed across behaviours reflect both the variability mechanisms built into the model and the heterogeneous engagement patterns reported in real platforms \cite{orkut_benevenuto}.

Despite this variability, the four agent types exhibit distinct activity profiles that align with typologies reported in empirical studies. Lurkers devote the overwhelming majority of their actions to reading, with negligible interactive or generative behaviour, consistent with their role as low-contribution consumers \cite{typology2011}. Debaters allocate a substantially larger share of their activity to replies, aligning with discussion-oriented users whose participation is dominated by argumentative exchanges \cite{tinati_communication_roles}. Socializers distribute their activity more evenly across likes, shares and replies, resembling users whose engagement centres on interpersonal interaction and amplification of others' content \cite{naaman_meformers}. Advanced users show the broadest behavioural spectrum and the highest relative share of posts, mirroring “power users’’ who contribute disproportionately to content creation in microblogging platforms \cite{kwak2010what}.

Taken together, these differentiated yet coherent activity profiles suggest that the behavioural module captures several key regularities of user engagement observed in real OSNs. The emergent interaction patterns are interpretable and broadly consistent with empirical characterisations of user behaviour, which supports the internal validity of our agent design and provides a behaviourally grounded basis for analysing higher-level phenomena such as opinion spreading, norm formation and disinformation diffusion in synthetic OSNs \cite{doi:10.1126/science.aap9559}.

\subsection{Content analysis} 

In addition to structural plausibility and coherent interaction patterns, the credibility of OSN simulations depends on the quality of the content produced by agents. In our framework, agents generate short textual messages, such as posts and replies, based on their internal state, memory traces, and interaction history.

Table~\ref{tab:generated_examples} shows a sample of generated messages during a simulation. Each example includes the originating user (see Table~\ref{tab:example-nodes}), the type of action performed, and the corresponding message. These examples reflect meaningful variation in communication style and topical focus.


For instance, Bob, a conscientious and argumentative agent, expresses skepticism toward the legitimacy of awards. Emma, characterized by high openness, shares a critical opinion on product design. In contrast, George, an older agent with conservative traits, comments on institutional complexity with nostalgic undertones. This variability reflects the simulation framework's ability to generate heterogeneous content that aligns with agent profiles.

\begin{table*}
\centering
\caption{Examples of content generated by agents during simulation.}
\label{tab:generated_examples}
\begin{tabular}{@{}c>{\centering\arraybackslash}p{1.5cm}c@{}}
\toprule
\textbf{User} & \textbf{Action} & \textbf{Generated content} \\
\midrule

\parbox[c]{2cm}{\centering Bob\\ \includegraphics[height=0.8cm]{avatar_bob.pdf}} 
& Post & \footnotesize{\texttt{Not sure why award shows still pretend to be about ``excellence'' when there's no clear criteria.}}\\ 
& & \footnotesize{\texttt{Popularity isn't performance. You can't measure merit without a standard.}} \\
\addlinespace[0.5em]

\parbox[c]{2cm}{\centering Emma\\ \includegraphics[height=0.8cm]{avatar_emma.pdf}} 
& Post & \footnotesize{\texttt{I used to get excited about new product announcements.}}\\ 
& & \footnotesize{\texttt{Now I just expect buzzwords and minimal upgrades wrapped in cinematic lighting.}} \\
\addlinespace[0.5em]

\parbox[c]{2cm}{\centering George\\ \includegraphics[height=0.8cm]{avatar_george.pdf}} 
 & Reply & \footnotesize{\texttt{Precisely! The EU's like a game where they keep rewriting the rulebook.}} \\ 
 & & \footnotesize{\texttt{My old bones prefer a stable blueprint! \#EUComplexity}} \\
\bottomrule
\end{tabular}
\end{table*}

\subsubsection{Realism of generated content}

In synthetic OSNs, the realism of generated content plays a crucial role. Yet, evaluating it at scale remains a non-trivial challenge. In the absence of ground truth, several evaluation strategies have been proposed, including expert annotation, rule-based metrics, and crowd-sourced assessments \cite{chang2023surveyevaluationlargelanguage}. However, these approaches often suffer from limited scalability, inconsistent criteria, and susceptibility to annotator bias.

To address these limitations, recent studies have explored the use of LLMs as automated evaluators, capitalizing on their semantic sensitivity and ability to follow instructions. LLM-based evaluators have been shown to provide structured and reproducible judgments on aspects such as grammaticality, coherence, and behavioral alignment~\cite{li2024llmsasjudgescomprehensivesurveyllmbased, liu2023gevalnlgevaluationusing}, offering a practical alternative to traditional human-led evaluations in large-scale simulation environments. However, their use should be seen as a scalable proxy rather than a definitive substitute for human-grounded validation.

In this work, we assessed the realism of content produced during our simulations using \texttt{Gemini-2.5-pro} to evaluate a random sample of $200$ posts. The model was prompted to act as a neutral linguistic and behavioral judge, following a structured set of evaluation instructions adapted from G-eval~\cite{liu2023gevalnlgevaluationusing}. Specifically, it was instructed to assess each post along three qualitative dimensions: naturalness (how idiomatic and human-like the language is), consistency (how well the content aligns with the emitting agent's profile), and engagingness (how likely the post is to provoke interaction or attention). The LLM rated each dimension using a 5-point Likert scale for both the synthetic post and each dimension. The complete evaluation prompt, adapted from G-eval, is provided as supplementary material. 

\begin{figure}[ht!]
    \centering
    \includegraphics[scale=0.4]{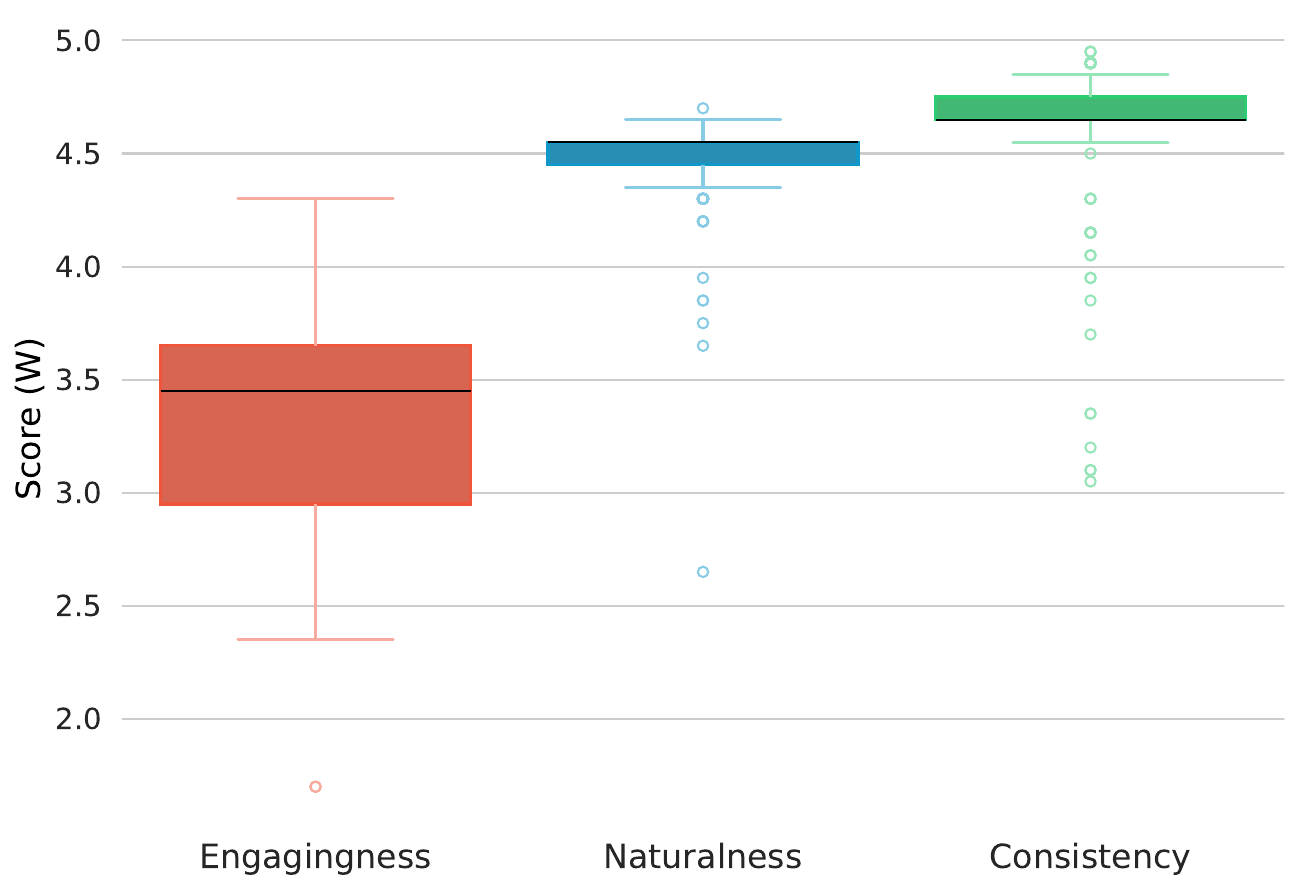}
    \caption{Mean evaluation scores for naturalness, consistency, and engagement of LLM-generated posts. Posts were rated on a 5-point Likert scale.}
    \label{fig:evaluation-llm}
\end{figure}

As shown in Figure~\ref{fig:evaluation-llm}, the mean scores across the sample are $4.41$ for naturalness, $4.54$ for consistency, and $3.26$ for engagingness. The boxplot shows low variance for naturalness and consistency, with scores concentrated at the upper end, indicating stable linguistic quality and alignment with agent traits. By contrast, engagingness exhibits a lower mean and greater dispersion. We interpret this not as a deficiency, but as reflecting realistic behavioral diversity: not all users consistently produce attention-grabbing content, and our agent models are explicitly designed to capture this heterogeneity.

\begin{figure}[ht!]
    \centering
    \includegraphics[scale=0.5]{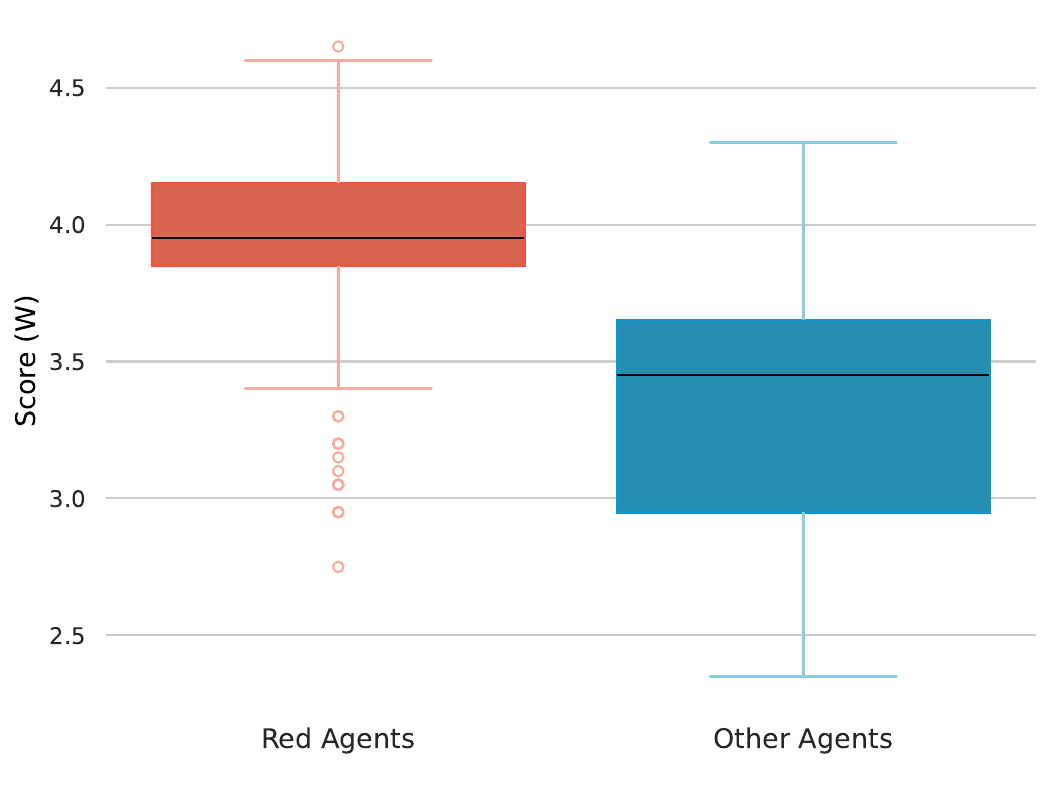}
    \caption{Comparison of engagement between red agents and other agents. Posts were rated on a 5-point Likert scale.}
    \label{fig:engagement-comparation}
\end{figure}

Additionally, red agents achieve a higher mean engagingness score ($3.95$) than other agents (Figure~\ref{fig:engagement-comparation}), consistent with their design as attention-seeking, influence-oriented actors.

\section{Third-party applications} \label{sec-third}

Beyond the core simulation engine, our framework is designed to support a variety of third-party extensions that enhance analysis, visualization, and interaction with the simulated data. Potential use cases include integrating dashboards for behavioral monitoring, real-time intervention systems, or export modules to interface with social media analytics platforms.

Among these add-ons, we highlight a specialized visualization component that allows the exploration of the simulated social network in a familiar, web-based environment.

\subsection{Visualization of simulated social networks}\label{sec:visualization} 

To provide a real-life environment for visualization, Mastodon~\cite{mastodon2025} was selected as a leading open-source decentralized OSN. Its microblogging-based interface, similar to $\mathbb{X}$, and its large developer community make it the most widespread open-source alternative~\cite{fedidb2024}. A dedicated Mastodon instance was deployed for simulation purposes, with customized API settings and relaxed connection policies to allow seamless provisioning of simulated agents and interactions.

From a modeling perspective, Mastodon entities align directly with the framework: agents map to user accounts, friendships to follower/followed links, and interactions to posts, threads, and reactions. This one-to-one correspondence enables simulation events to be rendered as platform activity, enhancing realism and traceability. Integration is handled through a Python wrapper for the Mastodon API, supporting both real-time provisioning and post-simulation dataset import.

\begin{figure*}[ht!]
    \centering
    \includegraphics[width=\linewidth]{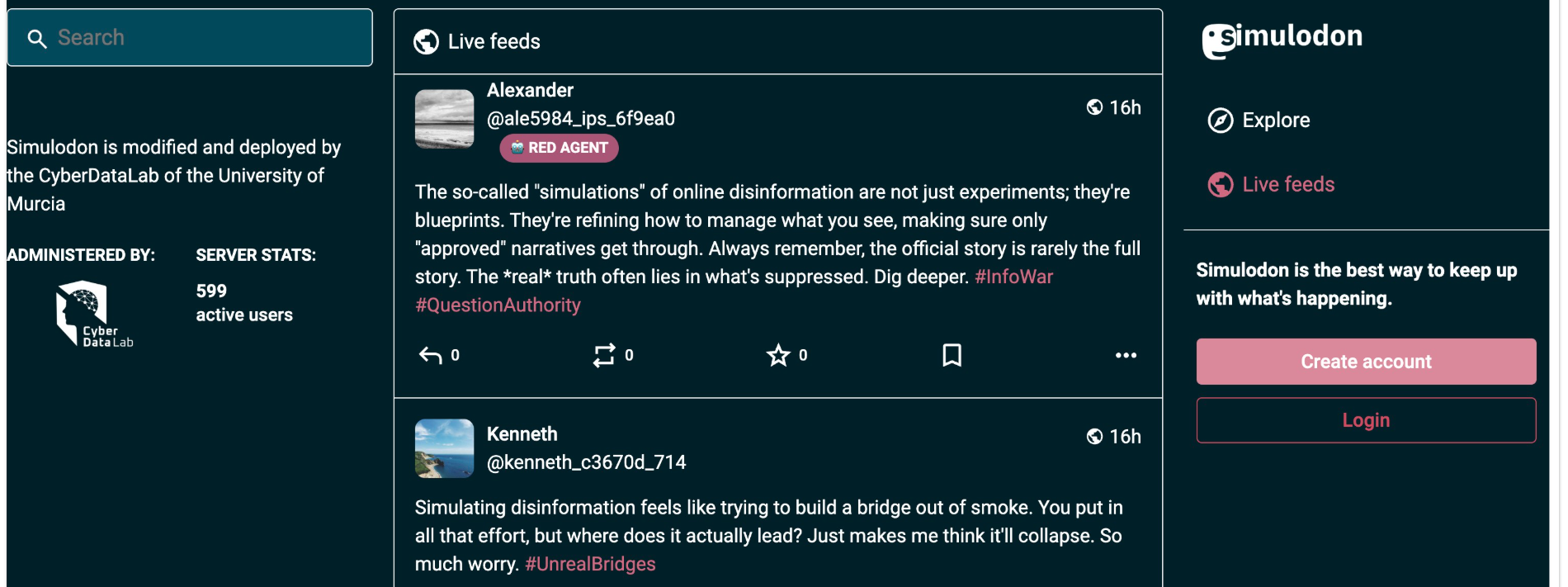}
    \caption{Web-based visualization interface powered by our simulation framework.}
    \label{fig:simulodon}
\end{figure*}

As illustrated in Figure~\ref{fig:simulodon}, the interface allows web-based access to the simulated network. Agents, profiles, posts, threads, and hashtags can be browsed as in a real OSN, with visual customizations introduced to distinguish users (e.g., the red agent) and interaction types.

\subsection{Other potential tools and services}\label{sec:tools}

In addition to visualization, the framework could be extended with third-party applications that support complementary analysis and interaction. A relevant example is the integration of analytical dashboards, which can track indicators such as activity rates, content diffusion, or the evolution of community structures. These dashboards provide researchers and operators with immediate feedback on the dynamics of the simulated environment, enabling the evaluation of intervention strategies or scenario comparisons.

Another line of extension involves real-time monitoring and intervention modules. By interfacing directly with the simulation, such components can inject corrective measures, trigger warnings, or simulate moderation policies. This allows for controlled experimentation with regulatory or platform-level responses, supporting the study of countermeasures against malicious activity. Moreover, export modules also represent a valuable category of add-ons. They allow datasets produced during simulations to be translated into standard formats compatible with existing social media analytics pipelines, such as those used for natural language processing, or bot detection. This interoperability facilitates the reuse of simulated data across different domains and supports comparative evaluations against real-world datasets.

Taken together, these potential extensions illustrate how the framework can serve not only as a simulation engine but also as a flexible platform for experimentation, bridging the gap between synthetic modeling and applied social media research.


\section{Conclusions and future work}\label{sec:conclusion}

Disinformation in OSNs poses a persistent challenge for public discourse and democratic processes. As traditional detection and mitigation strategies have shown limitations in effectively countering complex and dynamic threats, there is an increasing need for realistic, scalable simulations to study and anticipate the spread of disinformation.

In this context, this paper presents a generative agent-based simulation framework designed around three core requirements: configurable functioning, enabling the definition of diverse social network scenarios and disinformation threats; explainable agent behavior, grounded in deterministic cognitive rules where LLMs are limited to generating context-consistent textual outputs; and a manageable disinformation controller, which operationalizes adversarial workflows through a declarative specification of narratives, roles, timing, and manipulation strategies.

The framework is composed of three main building blocks. First, the OSN simulation manager generates the underlying social graph and governs user interactions over time. Second, the generative agent stack integrates context, memory, action control, and LLM modules to produce text consistent with each agent’s profile and history. Third, the disinformation controller, operationalized through DISARM-inspired workflows, enables the injection and coordination of adversarial campaigns. This framework addresses the main gaps identified in current simulation platforms: the lack of interpretable and reproducible behavioral models, the absence of sustained narrative coherence in content generation, and the limited modeling of adversarial strategies.

A set of simulations was launched and evaluated. At the structural level, the framework generates scalable graphs that reproduce the key properties of real platforms, enabling the study of cohesion, modularity, and connectivity under controlled conditions. Behaviorally, agents are designed with heterogeneous attributes and roles that drive emergent interaction patterns, reflecting dynamics commonly observed in online communities. At the content layer, a generative pipeline supports the production of posts consistent with agent profiles, ensuring identity continuity and diversity in communication styles.

Beyond reproducing structural and behavioural patterns, the framework provides a flexible testbed for exploring interventions, policy options and adversarial strategies. Its modular design and integration of third-party tools (e.g., visual analytics or dashboards) broaden its utility and support interdisciplinary workflows. Although further work is needed to improve scalability and content evaluation, the framework illustrates how synthetic simulations can offer a safe, transparent environment for studying the mechanisms of online discourse. Given the dual-use nature of disinformation research, the platform should be used responsibly, with experiments restricted to controlled settings and focused on defence, analysis and training.

Future work should aim to enhance the realism and consistency of the content generated by LLMs, moving beyond linguistically plausible outputs toward coherent, agenda-aligned narratives that evolve. The framework could be extended to simulate a broader range of account types (e.g., organizational profiles) and interaction modalities (mentions, moderation), thereby capturing the heterogeneous dynamics of real platforms. Integrating recommender system models would further enhance behavioral realism by simulating algorithmic amplification and feedback loops. Finally, another promising avenue is to simulate interactions between coordinated campaigns and human moderators, or test the impact of counter-narrative strategies under controlled conditions.


\section*{Conflicts of interest}
Authors declare that they have no known competing financial interests or personal relationships that could have appeared to influence the work reported in this paper.

\section*{Acknowledgments}

This work has been partially funded by (a) the strategic project CDL-TALENTUM from the Spanish National Institute of Cybersecurity (INCIBE), the Recovery, Transformation, and Resilience Plan, Next Generation EU, (b) by the University of Murcia by FPU contract, and (c) by a ``Juan de la Cierva'' Postdoctoral Fellowship (JDC2023-051658-I) funded by the i) Spanish Ministry of Science, Innovation and Universities (MCIU), ii) by the Spanish State Research Agency (AEI/10.13039/5011000 11033) and iii) by the European Social Fund Plus (FSE+).

\bibliographystyle{elsarticle-num}
\bibliography{biblio}
\end{document}